
\magnification=1200
\def\degg{{\rm deg}}
\def\deg'{{\rm deg}'}
\bigskip
\hfill{hep-th/9303152}
\vskip 3cm
\bigskip
\centerline{\bf DIFFERENTIAL GEOMETRY OF THE SPACE OF ORBITS}
\smallskip
\centerline{\bf OF A COXETER GROUP}
\medskip
\medskip
\centerline{\bf Boris DUBROVIN}
\smallskip
\centerline{\sl SISSA, via Beirut, 2}
\centerline{\sl I-34013 TRIESTE, Italy}
\centerline{{\sl E-mail} dubrovin@tsmi19.sissa.it}
\bigskip
\bigskip
{\bf Abstract.}
\medskip
Differential-geometric structures on the space of orbits of
a finite Coxeter group, determined by Groth\'endieck residues,
are calculated. This gives a construction of a 2D topological field theory
for an arbitrary Coxeter group.
\bigskip
\centerline{Ref. S.I.S.S.A. 29/93/FM (February 1993)}

\vfill\eject
{\bf Introduction. Formulation of main results.}
\medskip

Let $W$ be a (finite) Coxeter group, i.e. a finite group of linear
transformations
of a $n$-dimensional Euclidean space $V$ generated by reflections. The space
of orbits
$$ M = V/W$$
has a natural structure of affine variety: the coordinate ring of $M$
coincides with the ring $R := S(V)^W$ of $W$-invariant polynomial functions
on $V$. Due to Chevalley theorem this is a polynomial ring with the generators
$x_1$, ..., $x_n$ being invariant homogeneous polynomials. The basic invariant
polynomials are not specified uniquely. But their degrees $d_1$, ..., $d_n$
are invariants of the
group (see below Sect. 2). The maximal degree $h$ of the polynomials is
called Coxeter number of the group $W$. (More details about Coxeter groups
will be given in Sect. 2.)

A clue to understanding of a rich differential-geometric structure of the
orbit spaces can be found in the singularity theory. According to this
the
complexified
orbit space of an irreducible Coxeter group is bi-holomorphic equivalent to
the universal unfolding of a simple singularity [1, 10, 31, 41].
Under this identification the Coxeter group
coincides with the monodromy group of vanishing cycles
of the singularity. The discriminant of the Coxeter group (the set of
irregular orbits) is identified with the bifurcation diagram of the
singularity. The invariant Euclidean inner product on $V$ coincides
with the pairing on the cotangent bundle $T_*M$ defined by the intersection
form of vanishing cycles [45]. The bi-holomorphic equivalence is given
by the period
mapping.

Additional differential-geometric structures on
a universal unfolding of an isolated hypersurface singularity are determined
by the Groth\'endieck residues (see [39]). Let me explain
this for the simplest example of the group $A_n$
where the formulae for the residues were rediscovered by R.Dijkgraaf,
E. and H.Verlinde [15] (they also found new remarkable properties of these
residues, see below). This is obtained from the
group of all permutations of the coordinates $\xi_1$, ..., $\xi_{n+1}$ of
$(n+1)$-dimensional space by restriction onto the subspace
$$V = \{ \xi_1 + ... +\xi_{n+1} = 0\} .$$
The space of orbits of $A_n$ can be identified with the universal unfolding
of the simple singularity $f=z^{n+1}$,
$$M = \{ f(z;x_1, ..., x_n) = z^{n+1} + x_nz^{n-1}
+ ... + x_1 = \prod_{i=1}^{n+1} (z-\xi_i)\} .$$
The residue pairing defines a new metric on $M$: the inner product of two
tangent vectors in a point $x$ $= (x_1, ..., x_n)$ is defined by the formula
$$(\dot f(z; x(s_1)), \dot f(z;x(s_2)))_x := {\rm res}_{z=\infty}
{\dot f(z; x(s_1)) \dot f(z;x(s_2)) \over f'(z;x)}.
\eqno(1)$$
Here the dots mean derivatives w.r.t. the parameters $s_1$, $s_2$ resp. on two
curves through the point $x$, the prime means $d/dz$. This pairing does not
degenerate on $TM$. We can define in a similar way a trilinear form on
$TM$ putting
$$c(\dot f(z;x(s_1)), \dot f(z;x(s_2)), \dot f(z;x(s_3)))_x :=
{\rm res}_{z=\infty} {\dot f(z;x(s_1)) \dot f(z;x(s_2)) \dot f(z;x(s_3))
\over f'(z;x)}.
\eqno(2)$$
This gives rise to an operation of multiplication of tangent vectors
at any point $x\in M$
$$u,~v \mapsto u\cdot v, ~~u, v\in T_xM$$
uniquely specified by the equation
$$c(u,v,w)_x = (u\cdot v,w)_x .$$
This is a commutative associative algebra with a unity for any $x$ isomorphic
to the algebra of truncated polynomials
$${\bf C}[z]/(f'(z;x)).$$
At the origin $x=0$ the algebra coincides with the local algebra of the
$A_n$-singularity ${\bf C}[z]/(f'(z))$.

One can define in a similar way polylinear forms
$$c_k(u_1, \dots ,u_k)_x :=
{\rm res}_{z=\infty}{\dot f(z; x(s_1)) \dots \dot f(z;x(s_k))
\over f'(z;x)}$$
where the tangent vectors in the point $x$ have the form
$$u_i = \dot f(z;x(s_i)), i=1, \dots , k.$$
For $k>3$ they can be expressed in a pure algebraic way via the
multiplication of vectors and the pairing $(~,~)$:
$$c_k(u_1, \dots , u_k) = (u_1 \cdot u_2 \cdot \dots \cdot u_{k-1},
u_k).$$
Note that this formula coincides with the factorization rule for the
primary correlators in two-dimensional TFT [16, 50]. Further details about
2-dimensional topological
field theory from the point of view of the theory of singularities
can be found in [8].
\medskip

Let us come back to orbit spaces of arbitrary Coxeter groups. As it was
mentioned above the intersection form of the simple singularity
corresponding to a Coxeter group (as a metric
on the universal unfolding) on the space of orbits can be defined
intrinsicaly being induced by the invariant Euclidean structure in $V$.
V.I.Arnol'd in [3] formulated (for $A-D-E$-singularities;
for other simple singularities see [28])
the problem of calculation of the local algebra
structure in intrinsic terms, i.e. via the metric on the orbit space
$M$ (this metric was introduced by Arnol'd in [2]; it is
called also {\it convolution of invariants}). In the same time
K.Saito [36, 37] solved the problem of calculation
in intrinsic terms the residue pairing metric. The ideas of the
papers [3, 36, 37] (and of the paper [39] where the constructions of [36, 37]
were developed for extended affine root systems) are very important
for constructions of the present paper.

To develop the approaches of these works I am going to contribute to
understanding of the
problem of giving an intrinsic description of the
differential-geometric structures on the space of orbits of a Coxeter
group induced by the constructions of the theory of singularities.
[This problem was spelled out by K.Saito in [39] (but the structures
like (2) were not considered). He considered it as generalised Jacobi's
inversion problem: to describe the image of the period mapping.
An equivalent problem of axiomatization of the convolution of invariants
was formulated by Arnol'd in [5, p.72].]
I will give an intrinsic formula for calculation
of the Groth\'endieck residues for arbitrary Coxeter group without
using the construction of the correspondent universal unfolding.
My purpose is to obtain a complete differential-geometric
characterization of the space of orbits in terms of these structures
(see Conjecture at the end of this section).

I came to this problem from another side when I was trying to understand
a geometrical foundation of two-dimensional topological
field theories (TFT) [14-16, 42, 48 - 50]. The idea was to extend
the Atiyah's axioms [7] of TFT (for the two-dimensional case) by the
properties of the canonical moduli space of a TFT model proved in [15]
(see also [16]). On this way I found a nice geometrical object that I called
Frobenius manifold. Any model of two-dimensional TFT is encoded by a
Frobenius manifold and I showed that many constructions of TFT (integrable
hierarchies for the partition function, their bi-hamiltonian formalism
and $\tau$-functions, string equations, genus zero recursion relations
for correlators) can be deduced from geometry of Frobenius manifolds
[20, 22]. It looks like Frobenius manifolds play also an important r\^ole
in the theory of singularities. Better understanding of the r\^ole could
elucidate still misterious relations between the theory of singularities,
theory of integrable systems, and intersection theories on moduli spaces
of algebraic curves [12, 16, 29, 30, 48-52].

In the present paper I show that the orbit spaces of
Coxeter groups carry a natural structure of Frobenius manifold. For the
groups of $A-D-E$ series this gives an intrinsic description (i.e. only
in terms of the Coxeter group) of the residue structures like (1), (2)
(this coincides with the primary chiral algebra of the $A-D-E$-topological
minimal models [15, 42]).

It's time to proceed to the definition of Frobenius manifold. This is
a coordinate-free formulation of a differential equation arised
in [15, 49] (I called it WDVV - Witten-Dijkgraaf-E.Verlinde-H.Verlinde
equation). This is a system of equations for a function $F(t)$ of $n$
variables $t= (t^1, \dots ,t^n)$ resulting from the following conditions:

1. The matrix
$$\eta_{\alpha\beta} := {\partial^3 F(t)\over
\partial t^1 \partial t^\alpha \partial t^\beta}
$$
should be constant and
not degenerate. Let us denote by $(\eta^{\alpha\beta})$ the inverse matrix.

2. The coefficients
$$c_{\alpha\beta}^\gamma (t) := \sum_\epsilon \eta^{\gamma\epsilon}
{\partial^3 F(t)\over
\partial t^\epsilon \partial t^\alpha \partial t^\beta}$$
for any $t$ should be structure constants of an associative algebra.

3. The function $F(t)$ should be quasihomogeneous of a degree $3-d$
where the degree of the variables are $1-q_\alpha :=$ deg$\, t^\alpha$,
$q_1 = 0$. (In physical literature $d$ is called dimension of the TFT-model
and $q_\alpha$ are called charges of the primary fields.)

To give a coordinate-free reformulation of the WDVV equations let me recall
the notion of {\it Frobenius algebra}. This is a
commutative\footnote{$^\dagger$}{Also noncommutative Frobenius algebras
are considered by algebraists.} associative algebra with a unity over
a commutative
ring $R$ supplied with a symmetric nondegenerate $R$-bilinear inner
product $(~,~)$ being invariant\footnote{$^\ddagger$}{Invariant inner
product on a Frobenius algebra is not unique: any linear functional
$\omega \in A^*$ defines an invariant symmetric inner product on $A$ by
the formula $(a,b)_\omega := \omega (ab)$. This does not degenerate for
generic $\omega$. We consider a Frobenius algebra with a {\it marked}
invariant inner product.}
in the following sense
$$(ab,c) = (a,bc)$$
for any $a$, $b$, $c\in A$. We will consider the cases where $R =
{\bf R,~C}$ or the ring of functions on a manifold. There is a natural
operation of {\it rescaling} of a Frobenius algebra: for an invertible constant
$c$ we change the multiplication law, the unity $e$ and the invariant
inner product $(~,~)$ putting
$$a\cdot b \mapsto c a\cdot b, ~~ e\mapsto c^{-1} e, ~~(~,~)\mapsto
\varphi (c) (~,~)$$
for an arbitrary $\varphi (c)$.
\smallskip
{\bf Definition 1.} A manifold $M$ (real or complex) is called
{\it Frobenius manifold} if the tangent planes $T_xM$ are supplied with
a structure of Frobenius algebra smoothly depending on the point $x$
and satisfying the following properties.

1. The metric on $M$ specified by the invariant inner product $(~,~)$ is flat
(i.e. the curvature of the metric vanishes).

2. The unity vector field $e$ is covariantly constant
$$\nabla e = 0.$$
Here $\nabla$ is the Levi-Civit\`a connection for the metric $(~,~)$.

3. Let $c$ be the section of the bundle $S^3T_*M$ (i.e. a symmetric
trilinear form on $TM$) given by the formula
$$c(u,v,w) := (u\cdot v, w).$$
Then the tensor
$$(\nabla_zc)(u,v,w)$$
should be symmetric in $u$, $v$, $w$, $z$ for any vector fields
$u$, $v$, $w$, $z$.

4. A one-parameter group of diffeomorphisms should be defined on $M$
acting as rescalings on the algebras
$T_xM$.
\medskip
We will denote by $v$ the generator of the one-parameter group. It can be
normalised by the condition on the commutator of the vector fields $e$, $v$
$$[e,v] = e.$$
We will call $v$ {\it Euler vector field} on the Frobenius manifold. The
eigenvalues of the linear operator $\nabla_iv^j$ are called {\it invariant
degrees} of the Frobenius manifold.

In the flat coordinates $t^1$, ..., $t^n$ for $(~,~)$ the metric
is given by a constant matrix $(\eta_{\alpha\beta})$, the unity vector field
also has constant coordinates (we can normalize the flat coordinates in
such a way that $e = \partial / \partial t^1$) and the Euler vector field has
the form
$$v = \sum (1-q_\alpha )t^\alpha {\partial \over \partial t^\alpha}.$$
The degrees $(1-q_\alpha)$ of the coordinates $t^\alpha$ coincide with
the invariant degrees of the Frobenius manifold.
The tensor $c_{\alpha\beta\gamma}$ can be
represented (at least localy) as the third derivatives of a function
$F(t)$ satisfying WDVV equations.

The nondegenerate form $(~,~)$ establishes an isomorphism
$$(~,~) : TM \to T_*M.$$
This provides also the cotangent planes with a structure of Frobenius
algebra.

Note that the space of vector fields on a Frobenius manifold acquires a
natural structure of a Frobenius algebra over the ring $R$ of functions
on $M$. This can be used for algebraization of the notion of Frobenius
manifold for a suitable class of rings $R$ as a Frobenius $R$-algebra
structure on the $R$-module $Der\, R$ of derivations of $R$
satisfying the above properties [22]. Particularly, if $R = {\bf C}[x_1,
\dots , x_n]$ is a polynomial ring then a Frobenius $R$-algebra
structure on the  polynomial vector fields $Der\, R$
satisfying the conditions of Definition 1 will be called
{\it polynomial Frobenius manifold}
(see the algebraic reformulation of the notion of polynomial Frobenius
manifold in Appendix to this paper). In this case $M =$ Spec $R$
is an affine space and the correspondent solution of WDVV is a
quasihomogeneous polynomial. Polynomial solutions of WDVV with integer
and rational coefficients are of special interest due to their probable
relation to intersection theories on moduli spaces of algebraic curves
and their holomorphic maps [6, 52].

Let us come back to the orbit space $M$ of a Coxeter group. We denote
by $<~,~>^*$ the metric on the cotangent bundle $T_*M$ induced by the
$W$-invariant Euclidean structure on $V$. There are two marked vector fields
on $M$: the Euler vector field
$$E := \sum d_ix_i{\partial \over \partial x_i}$$
and the vector field
$$e := {\partial \over \partial x_1}$$
corresponding to the polynomial of the maximal degree deg $x_1 = h$. The vector
field $e$ is defined uniquely up to a constant factor. The {\it Saito metric}
on $M$ (inner product on $T_*M$) is defined as
$$(~,~)^* := {\cal L}_e <~,~>^*$$
(the Lie derivative along $e$). This is a flat globaly defined metric on $M$
[36, 37] (for convenience of the reader I reprove this statement in Sect. 2).
Our main technical observation inspired by the
differential-geometric theory of S.P.Novikov and the
author of Poisson brackets of hydrodynamic type [24, 25] and by bi-hamiltonian
formalism [33] is that any linear combination $a<~,~>^* + b(~,~)^*$
of the flat metrics is again flat.
\medskip
{\bf Theorem 1.} {\it There exists a unique (up to rescaling) polynomial
Frobenius
structure on the space of orbits of a finite irreducible Coxeter group with the
charges and dimension
$$q_\alpha = 1 - {d_\alpha\over h}, ~~d = 1-{2\over h},
\eqno(3)$$
the unity $e$, the Euler vector field ${1\over h}E$, and the Saito invariant
metric such that for any two invariant
polynomials $f$, $g$ the following formula holds
$$i_v(df\cdot dg) = <df, dg>^*.
\eqno(4)$$}

Here $i_v$ is the operator of inner derivative (contraction) along the vector
field $v$.

The formula (4) gives an effective method [23] for calculation of the structure
constants of the Frobenius manifold in the flat coordinates for the Saito
metric (see formula (2.25) below). If the Saito flat coordinates are chosen to
be invariant polynomials with rational coefficients then the polynomial $F(t)$
also has rational coefficients (it follows from (2.25)).

In the origin $t=0$ the structure constants $c_{ij}^k(0)$ of the Frobenius
algebra on $T_0M$ coincide with the structure constants of the local
algebra of the correspondent
simple singularity $F(z) = 0$
$$\phi_i(z) \phi_j(z) = c_{ij}^k(0) \phi_k(z)\, ({\rm mod}\, F'(z)).$$
Here $\phi_i(z) := \left[
\partial F(z; x_1, \dots , x_n) /\partial x_i \right]_{x=0}$, $F(z;
x_1, \dots ,x_n)$ is the versal deformation of the singularity $F(z) \equiv
F(z; 0) = 0$. In the origin the formula (4) thus coincides with the formula
of Arnol'd [3, 28] related the local algebra with the linearized convolution of
invariants (i.e., with the linear part of the Euclidean metric)
where the identification of $T_0M$ with the cotangent plane $T_{*0}M$
should be given by the Saito metric. But the formula
(4) gives more providing a possibility to calculate the local algebra
via the convolution of invariants.

\medskip
Let $R = {\bf C}[x_1, ...,x_n]$ be the coordinate ring of the orbit space $M$.
The Frobenius algebra structure on the tangent planes
$T_xM$ for any $x\in M$ provides the $R$-module $Der\, R$ of invariant
vector fields with a structure of Frobenius algebra over $R$.
To describe this structure
let us  consider such a basis of invariant polynomials $x_1$, ..., $x_n$
of the Coxeter group
that deg $x_1 = h$. Let $D(x_1, ..., x_n)$ be the discriminant of the group.
We introduce a polynomial of degree $n$ in an auxiliary variable $u$ putting
$$P(u; x_1, ..., x_n) := D(x_1 -u, x_2, ..., x_n).
\eqno(5)$$
Let $D_0(x_1, ..., x_n)$ be the discriminant of this polynomial in $u$.
It does not vanish identicaly on the space of orbits.
\medskip
{\bf Theorem 2.} {\it The map
$$1\mapsto e, ~~u\mapsto v
\eqno(6a)$$
can be extended uniquely to an isomorphism of $R$-algebras
$${\bf C}[u,x_1, \dots ,x_n]/(P(u;x)) \to Der\, R.
\eqno(6b)$$}
\medskip
{\bf Corollary.} {\it The algebra on $T_xM$ has no nilpotents
outside the zeroes of the polynomial $D_0(x_1, ..., x_n)$.}
\medskip
A non-nilpotent Frobenius algebra (over ${\bf C}$) can be decomposed
into a direct sum of one-dimensional Frobenius algebras. Warning:
by no means this implies even local decomposability of a massive (see
below) Frobenius manifold into a direct sum of one-dimensional Frobenius
manifolds.
\medskip
{\bf Definition 2.} A Frobenius manifold $M$ is called {\it massive}
if the algebra on the tangent planes $T_xM$ is non-nilpotent
for a generic $x\in M$.
\medskip
In physical language massive Frobenius manifolds correspond to massive
TFT models. Examples of massless TFT models where the algebra structure
on the tangent planes is identicaly nilpotent are given by topological
sigma-models with a Calabi-Yau target space [52].
\medskip
{\bf Conjecture.} {\it Any massive polynomial Frobenius manifold
with positive invariant degrees is isomorphic to the orbit space
of a finite Coxeter group.}
\medskip
In other words, the constructions of Theorem 1 (and their direct products)
give all massive polynomial
solutions of WDVV with
$$0\leq q_\alpha \leq d < 1.$$
This could give a simple approach to classification of 2-dimensional
topological
field theories with $d<1$. An alternative approach was developed recently by
S.Cecotti and C.Vafa [12]. It is based on studying of Hermitean metrics on
a Frobenius manifold obeying certain system of differential equations
(the so-called equation of topological-antitopological fusion [11], see also
[21]). The approach of [12] also gives rise to Coxeter groups (and their
generalizations) in classification of topological field theories.
\medskip
The Conjecture can be \lq\lq improved" a little: instead of polynomiality it
is sufficiently to assume that the function $F(t)$ is analytic in the
origin. For positive invariant degrees analyticity in the origin
implies polynomiality.

The Conjecture can be verified easily for  2- and 3-dimensional
manifolds. There are other strong evidences in support of the
conjecture. I am going to discuss them in a separate publication.
\medskip
{\bf Historical remark.} I started to think about polynomial solutions of
WDVV trying to answer a question of Vafa [43]: what are the 2-dimensional
topological field theories (in the approach based on WDVV equations) for
which the partition function is a power series in the coupling constants
with rational coefficients? The question was motivated by the interpretation,
due to E.Witten [42 - 44], of the logarythm of the partition function as a
generating
function of intersection numbers of cycles on some orbifolds. On this way I
found
the solutions (2.46) - (2.48); the sense of the solutions (2.47) and
(2.48) from the point of view of known topological field theories was not
clear.
In December 1992 during my talk at I.Newton institute Arnol'd immediately
recognized in the degrees of the polynomials (2.46) - (2.48) the Coxeter
numbers (plus one) of the three Coxeter groups in the three-dimensional space.
This became the starting point of the present work.

\medskip
\medskip
{\bf 1. Differential-geometric preliminaries.}
\medskip
The name {\it contravariant metric} (or, briefly, {\it
metric}) will mean a symmetric nondegenerate
bilinear form $<~,~>^{*}$ on the cotangent bundle $T_{*} M$
to a manifold $M$. In a local coordinate system $x^1$, ...,
$x^n$ the metric is given by its components
$$g^{ij} (x) := <dx^i,dx^j>^{*}
\eqno(1.1)$$
where $(g^{ij})$ is an invertible symmetric matrix. The inverse
matrix $(g_{ij}) := (g^{ij})^{-1}$ specifies a {\it covariant metric} $<~,~>$
on the manifold $M$ (usualy it is also called metric on the manifold) i.e. a
nondegenerate inner product on the tangent bundle
$TM$
$$<\partial_i,\partial_j> := g_{ij}(x)
\eqno(1.2)$$
$$\partial_i := {\partial \over \partial x^i}.
$$
The {\it Levi-Civit\`a connection} $\nabla_k$ for the metric is uniquely
specified by the conditions
$$\nabla_kg_{ij}: = \partial_k g_{ij} -\Gamma_{ki}^sg_{sj}
-\Gamma_{kj}^sg_{is} = 0
\eqno(1.3a)$$
or, equivalently,
$$\nabla_kg^{ij} := \partial_k g^{ij} +\Gamma_{ks}^i g^{sj}
+\Gamma_{ks}^j g^{is} =0
\eqno(1.3b)$$
and
$$\Gamma_{ij}^k = \Gamma_{ji}^k.
\eqno(1.4)$$
(Summation over twice repeated indices here and below is assumed.
We will keep the symbol of summation over more than twice repeated
indices.)
Here the coefficients $\Gamma_{ij}^k$ of the connection (the
Christoffel symbols) can be expressed via the metric and its
derivatives as
$$\Gamma_{ij}^k = {1\over 2}g^{ks}\left( \partial_ig_{sj}
+\partial_j g_{is} - \partial_s g_{ij}\right) .
\eqno(1.5)$$
For us it will be more convenient to work with the {\it contravariant
components} of the connection
$$\Gamma^{ij}_k := <dx^i, \nabla_k dx^j>^{*} = -g^{is}\Gamma_{sk}^j.
\eqno(1.6)$$
The equations (1.3) and (1.4) for the contravariant components read
$$\partial_k g^{ij} = \Gamma^{ij}_k +\Gamma^{ji}_k
\eqno(1.7)$$
$$g^{is}\Gamma_s^{jk} = g^{js}\Gamma_s^{ik}.
\eqno(1.8)$$
It is also convenient to introduce operators
$$\nabla^i = g^{is}\nabla_s
\eqno(1.9a)$$
$$\nabla^i\xi_k = g^{is}\partial_s\xi_k + \Gamma^{is}_k\xi_s.
\eqno(1.9b)$$
For brevity we will call the operators $\nabla^i$ and the correspondent
coefficients $\Gamma_k^{ij}$ {\it contravariant connection}.

The {\it curvature tensor} $R_{slt}^k$ of the metric measures noncommutativity
of the operators $\nabla_i$ or, equivalently $\nabla^i$
$$(\nabla_s\nabla_l -\nabla_l\nabla_s)\xi_t = -R_{slt}^k\xi_k
\eqno(1.10a)$$
where
$$R_{slt}^k = \partial_s\Gamma_{lt}^k -\partial_l\Gamma_{st}^k
+\Gamma_{sr}^k \Gamma_{lt}^r -\Gamma_{lr}^k \Gamma_{st}^r.
\eqno(1.10b)$$
We say that the metric is {\it flat} if the curvature of it vanishes.
For a flat metric local {\it flat coordinates} $p^1$, ..., $p^n$ exist
such that in these coordinates the metric is constant and the components
of the Levi-Civit\`a connection vanish. Conversely, if a system of flat
coordinates for a metric exists then the metric is flat. The flat coordinates
are determined uniquely up to an affine  transformation with constant
coefficients. They
can be found from the following system
$$\nabla^i\xi_j = g^{is}\partial_s\partial_jp + \Gamma^{is}_j\partial_sp = 0,
{}~i,j = 1,...,n.\eqno(1.11)$$
If we choose the flat coordinates orthonormalized
$$<dp^a,dp^b>^{*} =\delta^{ab}
\eqno(1.12)$$
then for the components of the metric and of the Levi-Civit\`a connection
the following formulae hold
$$g^{ij} = {\partial x^i\over \partial p^a}{\partial x^j\over \partial p^a}
\eqno(1.13a)$$
$$\Gamma^{ij}_kdx^k = {\partial x^i\over \partial p^a}
{\partial^2 x^j\over \partial p^a \partial p^b}dp^b.
\eqno(1.13b)$$

All these facts are standard in geometry (see, e.g., [26]). We need
to represent the formula (1.10b) for the curvature tensor in a slightly
modified form (cf. [25, formula (2.18)]).
\medskip
{\bf Lemma 1.1.} {\it For the curvature of a metric the following formula holds
$$R^{ijk}_l := g^{is}g^{jt} R_{slt}^k =
g^{is}\left( \partial_s\Gamma_l^{jk} - \partial_l\Gamma_s^{jk}\right)
+\Gamma_s^{ij}\Gamma_l^{sk} - \Gamma_s^{ik}\Gamma_l^{sj}.
\eqno(1.14)$$
}

Proof. Multiplying the formula (1.10b) by  $g^{is}g^{jt}$ and using (1.6) and
(1.7) we obtain (1.14). The lemma is proved.
\medskip
Let us consider now a manifold supplied with two nonproportional metrics
$<~,~>^{*}_1$ and $<~,~>^{*}_2$. In a coordinate system they are given by
their components $g^{ij}_1$ and $g^{ij}_2$ resp. I will denote by
$\Gamma_{1k}^{ij}$ and $\Gamma_{2k}^{ij}$ the correspondent Levi-Civit\`a
connections $\nabla_1^i$ and $\nabla_2^i$. Note that the difference
$$\Delta^{ijk} =
g_2^{is} \Gamma_{1s}^{jk} - g_1^{is} \Gamma_{2s}^{jk}
\eqno (1.15)$$
is a tensor on the manifold.
\medskip
{\bf Definition 1.1.} We say that the two metrics form a {\it flat pencil}
if:

1. The metric
$$g^{ij} = g_1^{ij} + \lambda g_2^{ij}
\eqno(1.16a)$$
is flat for arbitrary $\lambda$ and

2. The Levi-Civit\`a connection for the metric (1.16a) has the form
$$\Gamma_k^{ij} = \Gamma_{1k}^{ij} + \lambda \Gamma_{2k}^{ij}.
\eqno(1.16b)$$
\medskip
{\bf Proposition 1.1.} {\it For a flat pencil of metrics a vector field
$f = f^i\partial_i$ exists
such that the difference tensor (1.15) and the metric $g_1^{ij}$ have the
form
$$\Delta^{ijk} = \nabla_{2}^i\nabla_{2}^jf^k
\eqno(1.17a)$$
$$g_1^{ij} = \nabla_{2}^if^j + \nabla_{2}^jf^i + c g_2^{ij}
\eqno(1.17b)$$
for a constant $c$.
The vector field should satisfy the equations
$$\Delta_s^{ij} \Delta_l^{sk} = \Delta_s^{ik}\Delta_l^{sj}
\eqno(1.18)$$
where
$$\Delta_k^{ij} := g_{2ks} \Delta^{sij} = \nabla_{2k} \nabla_2^i f^j,$$
$$(g_1^{is}g_2^{jt} - g_2^{is}g_1^{jt} ) \nabla_{2s}\nabla_{2t}f^k = 0.
\eqno(1.19)$$
Conversely, for a flat metric $g_2^{ij}$ and for a solution $f$ of the system
(1.18), (1.19) the metrics $g_2^{ij}$ and (1.17b) form a flat pencil.}

Proof. Let us assume that $x^1$, ..., $x^n$ is the flat coordinate
system for the metric $g_2^{ij}$. In these coordinates we have
$$\Gamma_{2k}^{ij} = 0, ~ \Delta_k^{ij}
:= g_{2ks} \Delta^{sij} = \Gamma_{1k}^{ij} .
\eqno(1.20)$$
The equation $R^{ijk}_l = 0$ in these coordinates reads
$$\left( g_1^{is} + \lambda g_2^{is}\right)
\left( \partial_s\Delta_l^{jk} - \partial_l\Delta_s^{jk}\right)
+ \Delta_s^{ij} \Delta_l^{sk} - \Delta_s^{ik} \Delta_l^{sj} =0.
\eqno(1.21)$$
Vanishing of the linear in $\lambda$ term provides existence of a
tensor $f^{ij}$ such that
$$\Delta_k^{ij} = \partial_kf^{ij}.$$
The rest part of (1.21) gives (1.18). Let us use now the condition
of symmetry (1.8) of the connection (1.16b). In the coordinate system this
reads
$$\left( g_1^{is} +\lambda g_2^{is}\right) \partial_s f^{jk} =
\left( g_1^{js} +\lambda g_2^{js}\right) \partial_s f^{ik}.
\eqno(1.22)$$
Vanishing of the terms in (1.22) linear in $\lambda$ provides existence
of a vector field $f$ such that
$$f^{ij} = g_2^{is}\partial_s f^j.$$
This implies (1.17a).
The rest part of the equation (1.22) gives (1.19). The last equation (1.7)
gives (1.17b). The first part of the proposition is proved. The converse
statement follows from the same equations.
\medskip
{\bf Remark.} The theory of S.P.Novikov and the author establishes a
one-to-one correspondence between flat contravariant metrics on a manifold
$M$ and Poisson brackets of hydrodynamic type on the loop space
$$L(M) := \{ {\rm smooth ~maps} ~ S^1 \to M\} $$
with certain nondegeneracy conditions [24, 25]. For a flat metric $g^{ij}(x)$
and the correspondent contravariant connection $\nabla^i$ the Poisson bracket
of two functionals of the form
$$I = I[x] = {1\over 2\pi} \int_0^{2\pi} P(s,x(s))\, ds, ~~
J = J[x] = {1\over 2\pi} \int_0^{2\pi} Q(s,x(s))\, ds,
$$
$x = (x^i(s))$, $x(s+2\pi )
= x(s)$ is defined by the formula
$$\{ I,J\} := {1\over 2\pi }\int_0^{2\pi} {\delta I\over \delta x^i(s)}
\nabla^i{\delta J\over \delta x^j(s)} \, dx^j(s) +
{1\over 2\pi }\int_0^{2\pi} {\delta I\over \delta x^i(s)}
g^{ij}(x) d_s{\delta J\over \delta x^j(s)}.
$$
Here the variational derivative $\delta I / \delta x^i(s) \in
T_*M|_{x=x(s)}$ is defined by the equality
$$I[x+\delta x] - I[x] =
{1\over 2\pi}\int_0^{2\pi} {\delta I\over \delta x^i(s)} \delta
x^i(s)\, ds + o(|\delta x|);$$
$\delta J / \delta x^j(s) $ is defined by the same formula,
$d_s := ds {\partial\over \partial s}$. The Poisson bracket can be uniquely
extended
to all \lq\lq good" functionals on the loop space by Leibnitz rule [24, 25].
Flat pencils of metrics correspond to compatible pairs of Poisson
brackets of hydrodynamic type.
By the definition, Poisson brackets $\{ ~,~\}_1$ and
$\{ ~,~\}_2$ are called compatible if an arbitrary linear combination
$$a\{ ~,~\}_1 + b\{ ~,~\}_2$$
again is a Poisson bracket. Compatible pairs of Poisson brackets are
important in the theory of integrable systems [33].
\medskip

The main source of flat pencils is provided by the following statement.
\medskip
{\bf Lemma 1.2.} {\it If for a flat metric  in some coordinate
system $x^1$, ..., $x^n$ both the components $g^{ij}(x)$ of the metric and
$\Gamma_k^{ij}(x)$ of the correspondent Levi-Civit\`a connection depend
linearly
on the coordinate $x^1$ then the metrics
$$g_1^{ij} := g^{ij} ~ {\rm and }~ g_2^{ij} := \partial_1 g^{ij}
\eqno(1.23)$$
form a flat pencil if ${\rm det} (g_2^{ij}) \neq 0$. The correspondent
Levi-Civit\`a connections have the form
$$\Gamma_{1k}^{ij} := \Gamma_k^{ij}, ~\Gamma_{2k}^{ij} :=
\partial_1 \Gamma_k^{ij}.
\eqno(1.24)$$
}

Proof. The equations (1.7), (1.8) and the equation of vanishing of the
curvature have constant coefficients. Hence the transformation
$$g^{ij}(x^1, ..., x^n)\mapsto g^{ij}(x^1 +\lambda , ...,
x^n), ~\Gamma_k^{ij}(x^1, ..., x^n)\mapsto
\Gamma_k^{ij}(x^1 +\lambda , ..., x^n)$$
for an arbitrary $\lambda$ maps the solutions of these equations to
the solutions. By the assumption we have
$$g^{ij}(x^1+\lambda , ..., x^n) = g_1^{ij}(x) +\lambda g_2^{ij}(x),
{}~\Gamma_k^{ij}(x^1+\lambda , ..., x^n)
=\Gamma_{1k}^{ij}(x) +\lambda \Gamma_{2k}^{ij}(x).$$
The lemma is proved.
\medskip
All the above considerations can be applied also to complex (analytic)
manifolds where the metrics are quadratic forms analyticaly depending
on the point of $M$.
\medskip
\medskip
{\bf 2. Frobenius structure on the space of orbits of a Coxeter group.}
\medskip
Let $W$ be a {\it Coxeter group}, i.e. a finite group of linear transformations
of real $n$-dimensional space $V$ generated by reflections. We always
can assume the transformations of the group to be orthogonal w.r.t. a Euclidean
structure on $V$. The complete classification of irreducible Coxeter
groups was obtained in [13]; see also [9]. The complete list consists
of the groups (dimension of the space $V$ equals the subscript in the name
of the group) $A_n$, $B_n$, $D_n$, $E_6$, $E_7$, $E_8$, $F_4$, $G_2$
(the Weyl groups of the correspondent simple Lie algebras), the groups
$H_3$ and $H_4$ of symmetries of the regular
icosahedron and of the regular 600-cell in the 4-dimensional space
resp. and the groups $I_2(k)$ of symmetries of
the regular $k$-gone on the plane. The group $W$ also acts on the symmetric
algebra $S(V)$ (polynomials of the coordinates of $V$) and on the $S(V)$-module
$\Omega (V)$ of differential forms on $V$ with polynomial coefficients.
The subring $R=S(V)^W$ of $W$-invariant polynomials is generated by $n$
algebraicaly independent homogeneous polynomials $x^1$, ..., $x^n$ [9].
The submodule $\Omega(V)^W$ of the $W$-invariant differential forms with
polynomial coefficients is a free $R$-module with the basis
$dx^{i_1}\wedge ...\wedge dx^{i_k}$ [9]. Degrees of the basic invariant
polynomials are uniquely defined by the Coxeter group. They can be expressed
via the {\it exponents} $m_1$, ..., $m_n$ of the group, i.e. via the
eigenvalues
of a Coxeter element $C$ in $W$ [9]
$$d_i := {\rm deg}\, x^i = m_{n-i+1} +1,
\eqno(2.1a)$$
$$\{ {\rm eigen}~ C \} = \{ \exp {2\pi i (d_1 -1)\over h}, ...,
\exp {2\pi i (d_n -1)\over h}\} .
\eqno(2.1b)$$
The maximal degree $h$ is called {\it Coxeter number} of $W$. I will use the
reversed ordering of the invariant polynomials
$$d_1 =h >d_2 \geq ...\geq d_{n-1} >d_2=2.
\eqno(2.2)$$
The degrees satisfy the {\it duality condition}
$$d_i +d_{n-i+1} = h+2, ~i=1, ..., n.
\eqno(2.3)$$
The list of the degrees for all the Coxeter groups is given in Table 1.
\medskip
$$\matrix{W & d_1, ~..., ~d_n \cr
{} & {} \cr
A_n & d_i =n+2-i \cr
B_n & d_i = 2(n-i+1) \cr
D_n,~ n=2k & d_i=2(n-i), ~i\leq k, \cr
{} & d_i=2(n-i+1), ~k+1\leq i \cr
D_n, ~n=2k+1 & d_i=2(n-i), ~i\leq k, \cr
{} & ~d_{k+1} = 2k+1, \cr
{} & ~ d_i=2(n-i+1),
{}~k+2\leq i \cr
E_6 & 12,~9,~8,~6,~5,~2 \cr
E_7 & 18,~14,~12,~10,~8,~6,~2 \cr
E_8 & 30,~24,~20,~18,~14,~12,~8,~2 \cr
F_4 & 12,~8,~6,~2 \cr
G_2 & 6,~2 \cr
H_3 & 10,~6,~2 \cr
H_4 & 30,~20,~12,~2 \cr
I_2(k) & k,~2}$$
\smallskip
\centerline{ Table 1.}
\medskip
\medskip
I will extend the action of the group $W$ to the complexified space $V\otimes
{\bf C}$. The space of orbits
$$M = V\otimes {\bf C}/W$$
has a natural structure of an affine algebraic variety: the coordinate ring
of $M$ is the (complexified) algebra $R$ of
invariant polynomials of the group $W$.
The coordinates $x^1$, ..., $x^n$ on $M$ are defined up to an invertible
transformation
$$x^i\mapsto {x^i}'(x^1,...,x^n),
\eqno(2.4)$$
where ${x^i}'(x^1,...,x^n)$ is a graded homogeneous polynomial of the same
degree
$d_i$ in the variables $x^1$, ..., $x^n$, ${\rm deg}\, x^k = d_k$. Note that
the
Jacobian ${\rm det}(\partial {x^i}'/\partial x^j)$ is a constant (it should not
be zero). The transformations (2.4) leave invariant the vector field
$\partial_1 := \partial /\partial x^1$ (up to a constant factor) due to the
strict inequality $d_1 > d_2$. The coordinate $x^n$ is determined uniquely
within a factor. Also the vector field
$$E = d_1x^1\partial_1 + ... +d_nx^n\partial_n =
p^a {\partial \over \partial p^a}
\eqno(2.5)$$
(the generator of scaling transformations) is well-defined on $M$.

Let $<~,~>$ denotes the $W$-invariant Euclidean metric in the space $V$.
I will use the orthonormal coordinates $p^1$, ..., $p^n$ in $V$ with respect to
this metric. The invariant $x^n$ can be chosen as
$$x^n = {1\over 2} ((p^1)^2 + ... + (p^n)^2).
\eqno(2.6)$$
We extend $<~,~>$ onto $V\otimes {\bf C}$ as a complex quadratic form.

The factorization map $V\otimes {\bf C} \to M$ is a local diffeomorphism
on an open subset of $V\otimes {\bf C}$. The image of this subset in $M$
consists of {\it regular orbits} (i.e. the number of points of the orbit
equals $\#\,  W$). The complement is the {\it discriminant} ${\rm Discr} W$.
By the definition it consists of all irregular orbits. Note that the linear
coordinates in $V$ can serve also as local coordinates in small
domains in $M\setminus {\rm Discr} W$. It defines a metric $<~,~>$
(and $<~,>^*$) on $M\setminus {\rm Discr} W$. The contravariant metric
can be extended onto $M$ according to the following statement
(cf. [39, Sections 5 and 6]).

\medskip
{\bf Lemma 2.1.} {\it The Euclidean metric
of $V$ induces polynomial contravariant
metric $<~,~>^*$ on the space of orbits
$$g^{ij}(x) = <dx^i,dx^j>^* := {\partial x^i\over \partial p^a}
{\partial x^j\over \partial p^a}
\eqno(2.7)$$
and the correspondent contravariant Levi-Civit\`a connection
$$\Gamma_k^{ij}(x)dx^k = {\partial x^i\over \partial p^a}
{\partial^2x^j\over \partial p^a \partial p^b} dp^b
\eqno(2.8)$$
also is a polynomial one.}

Proof. The right-hand sides in (2.7)/(2.8)  are $W$-invariant
polynomials/differential forms with polynomial coefficients. Hence
$g^{ij}(x)$/$\Gamma_k^{ij}(x)$ are polynomials in $x^1$, ..., $x^n$.
Lemma is proved.
\medskip
{\bf Remark.} The matrix $g^{ij}(x)$ does not degenerate on
$M\setminus {\rm Discr} W$ where the factorization
$V\otimes {\bf C} \to M$ is a local diffeomorphism. So the polynomial
(also called {\it discriminant} of $W$)
$$D(x) := {\rm det} (g^{ij}(x))
\eqno(2.9)$$
vanishes precisely on the discriminant ${\rm Discr}W$ where the variables
$p^1$, ..., $p^n$ fail to be local coordinates. Due to this fact the matrix
$g^{ij}(x)$ often is called {\it discriminant matrix} of $W$.
The operation $x^i, x^j \mapsto g^{ij}(x)$ is also called {\it convolution
of invariants} (see [2]). Note that the image of $V$ in the real
part of $M$ is specified by the condition of positive semidefiniteness
of the matrix $(g^{ij}(x))$ (cf. [34]). The Euclidean connection
(2.8) on the space of orbits is called {\it Gauss - Manin connection}.
\medskip
{\bf Corollary 2.1.} {\it The functions $g^{ij}(x)$ and $\Gamma_k^{ij}(x)$
depend linearly on $x^1$.}

Proof. From the definition one has that $g^{ij}(x)$ and $\Gamma_k^{ij}(x)$
are graded homogeneous polynomials of the degrees
$${\rm deg}\, g^{ij}(x) = d_i + d_j -2
\eqno(2.10)$$
$${\rm deg}\,\Gamma_k^{ij}(x) = d_i + d_j - d_k - 2.
\eqno(2.11)$$
Since $d_i +d_j \leq 2h = 2d_1$ these polynomials can be at most linear
in $x^1$. Corollary is proved.
\medskip
{\bf Corollary 2.2} (K.Saito) {\it The matrix
$$\eta^{ij}(x) := \partial_1 g^{ij}(x)
\eqno(2.12)$$
has a triangular form
$$\eta^{ij}(x) = 0 ~{\rm for}~ i+j > n+1,
\eqno(2.13)$$
and the antidiagonal elements
$$\eta^{i(n-i+1)} =: c_i
\eqno(2.14)$$
are nonzero constants. Particularly,
$$c := {\rm det} (\eta^{ij}) = (-1)^{n(n-1)\over 2} c_1...c_n \neq 0.
\eqno(2.15)$$}

Proof. One has
$${\rm deg}\, \eta^{ij}(x) = d_i +d_j - 2 - h.$$
Hence deg $\eta^{i(n-i+1)} = 0$ (see (2.3)) and deg $\eta^{ij}<0$
for $i+j >n+1$. This proves triangularity of the matrix and
constancy of the antidiagonal entries $c_i$. To prove
nondegenerateness of $(\eta^{ij}(x))$ we consider, following
Saito, the discriminant (2.9) as a polynomial in $x^1$
$$D(x) = c(x^1)^n + a_1(x^1)^{n-1} + ... + a_n$$
where the coefficients $a_1$, ..., $a_n$ are quasihomogeneous
polynomials in
$x^2$, ..., $x^n$ of the degrees $h$, ..., $nh$ resp.
and the leading coefficient $c$ is given in (2.15).
Let $\gamma$ be the eigenvector of a Coxeter transformation $C$
with the eigenvalue $\exp (2\pi i/h)$. Then
$$x^k(\gamma ) = x^k(C\gamma) = x^k(\exp (2\pi i/h) \gamma ) =
\exp (2\pi id_k/h)x^k(\gamma).$$
For $k>1$ we obtain
$$x^k(\gamma) = 0,~ k=2,...,n.$$
But $D(\gamma)\neq 0$ [9]. Hence the leading coefficient $c\neq 0$.
Corollary is proved.
\medskip
{\bf Corollary 2.3.} {\it The space $M$ of orbits of a finite Coxeter
group carries a flat pencil of metrics $g^{ij}(x)$ (2.7) and $\eta^{ij}(x)$
(2.12) where the matrix $\eta^{ij}(x)$ is polynomialy invertible
globaly on $M$.}
\medskip
We will call (2.12) {\it Saito metric} on the space of orbits. This metric
will be denoted by $(~,~)^*$ (and by $(~,~)$ if considered on the tangent
bundle $TM$).
Let us denote by
$$\gamma^{ij}_k(x) := \partial_1\Gamma_k^{ij}(x)
\eqno(2.16)$$
the components of the Levi-Civit\`a connection for the metric $\eta^{ij}(x)$.
These are quasihomogeneous polynomials of the degrees
$${\rm deg}\,\gamma_k^{ij}(x) = d_i+d_j-d_k-h-2.
\eqno(2.17)$$
\medskip
{\bf Corollary 2.4} (K.Saito). {\it There exist homogeneous polynomials
$t^1(p)$, ..., $t^n(p)$ of degrees $d_1$, ..., $d_n$ resp. such that
the matrix
$$\eta^{\alpha\beta} := \partial_1 <dt^\alpha ,dt^\beta >^*
\eqno(2.18)$$
is constant.}

The coordinates $t^1$, ..., $t^n$ on the orbit space
will be called {\it Saito flat coordinates}. They can be chosen in such a
way that the matrix (2.18) is antidiagonal
$$\eta^{\alpha\beta} = \delta^{\alpha + \beta , n+1}.$$
Then the Saito flat coordinates are defined uniquely up to an $\eta$-orthogonal
transformation
$$t^\alpha \mapsto a^\alpha_\beta t^\beta ,$$
$$\sum_{\lambda + \mu = n+1} a^\alpha_\lambda a^\beta_\mu =
\delta^{\alpha + \beta , n+1}.$$

Proof. From flatness of the metric $\eta^{ij}(x)$ it follows that the flat
coordinates $t^\alpha (x)$, $\alpha = 1$, ..., $n$ exist at least localy.
They are to be determined from the following system
$$\eta^{is}\partial_s\partial_j t + \gamma_j^{is}\partial_st = 0
\eqno(2.19)$$
(see (1.11)). The inverse matrix $(\eta_{ij}(x)) = (\eta^{ij}(x))^{-1}$
also is polynomial in $x^1$, ..., $x^n$. So rewriting the system (2.19)
in the form
$$\partial_k\partial_l t + \eta_{il}\gamma_k^{is}\partial_st = 0
\eqno(2.20)$$
we again obtain a system with polynomial coefficients. It can be written as a
first-order system for the entries $\xi_l = \partial_lt$,
$$\partial_k\xi_l  + \eta_{il}\gamma_k^{is}\xi_s = 0, ~
k,l = 1,...,n
\eqno(2.21)$$
(the integrability condition $\partial_k\xi_l = \partial_l\xi_k$
follows from (1.4)). This is an overdetermined completely integrable system.
So the space of solutions has dimension $n$. We can choose a fundamental
system of solutions $\xi_l^\alpha (x)$ such that $\xi_l^\alpha (0) =
\delta_l^\alpha$. These functions are analytic in $x$ for sufficiently
small $x$.
We put $\xi_l^\alpha (x) =: \partial_lt^\alpha (x)$, $t^\alpha (0) = 0$.
The system of solutions is invariant w.r.t. the scaling transformations
$$x^i \mapsto c^{d_i} x^i,~ i=1, ..., n.$$
So the functions $t^\alpha (x)$ are quasihomogeneous in $x$ of the same degrees
$d_1$, ..., $d_n$. Since all the degrees are positive the power series
$t^\alpha (x)$ should be polynomials in $x^1$, ..., $x^n$. Because of the
invertibility of the transformation $x^i\mapsto t^\alpha$ we conclude that
$t^\alpha(x(p))$ are polynomials in $p^1$, ..., $p^n$. Corollary is proved.
\medskip
We need to calculate particular components of the metric $g^{\alpha\beta}$
and of the correspondent Levi-Civit\`a connection in the coordinates
$t^1$, ..., $t^n$ (in fact, in arbitrary homogeneous coordinates
$x^1$, ..., $x^n$).
\medskip
{\bf Lemma 2.2.} {\it Let the coordinate $t^n$ be normalized as in (2.6).
Then the following formulae hold:
$$g^{n\alpha} = d_\alpha t^\alpha
\eqno(2.22)$$
$$\Gamma_\beta^{n\alpha} = (d_\alpha -1) \delta_\beta^\alpha .
\eqno(2.23)$$}

(In the formulae there is no summation over the repeated indices!)

Proof. We have
$$g^{n\alpha} = {\partial t^n\over \partial p^a}
{\partial t^\alpha \over \partial p^a} =
p^a {\partial t^\alpha\over \partial p^a} = d_\alpha t^\alpha$$
due to the Euler identity for the homogeneous functions $t^\alpha (p)$.
Furthermore,
$$\Gamma_\beta^{n\alpha} dt^\beta = {\partial t^n\over \partial p^a}
{\partial^2 t^\alpha \over \partial p^a\partial p^b} dp^b =
p^a {\partial^2t^\alpha\over \partial p^a\partial p^b} dp^b =
p^a d\left( {\partial t^\alpha \over \partial p^a}\right) =
$$
$$d \left( p^a  {\partial t^\alpha \over p^a}\right) -
{\partial t^\alpha \over \partial p^a} dp^a =
(d_\alpha -1) dt^\alpha .$$
Lemma is proved.
\medskip
We can formulate now the main result of this section.
\medskip
{\bf Main lemma.} {\it Let $t^1$, ..., $t^n$ be the Saito flat coordinates
on the space of orbits of a finite Coxeter group and
$$\eta^{\alpha\beta} = \partial_1 <dt^\alpha, dt^\beta >^*
\eqno(2.24)$$
be the correspondent constant Saito metric.
Then there exists a
quasihomogeneous polynomial $F(t)$ of the degree $2h+2$ such that
$$<dt^\alpha, dt^\beta >^* = {(d_\alpha +d_\beta -2)\over h}
\eta^{\alpha\lambda}
\eta^{\beta\mu}\partial_\lambda\partial\mu F(t).
\eqno(2.25)$$
The polynomial $F(t)$ determines on the space of orbits a polynomial Frobenius
structure with the structure constants
$$c_{\alpha\beta}^\gamma (t) = \eta^{\gamma\epsilon}
\partial_\alpha\partial_\beta\partial_\epsilon F(t)
\eqno(2.26a)$$
the unity
$$e = \partial_1
\eqno(2.26b)$$
and the invariant inner product $\eta$.}

Proof. Because of Corollary 2.3 in the flat coordinates the tensor
$\Delta_\gamma^{\alpha\beta} = \Gamma_\gamma^{\alpha\beta}$ should
satisfy the equations (1.17) - (1.19) where $g_1^{\alpha\beta}
= g^{\alpha\beta}(t)$, $g_2^{\alpha\beta} = \eta^{\alpha\beta}$.
First of all according to (1.17a) we can represent the tensor
$\Gamma_\gamma^{\alpha\beta}(t)$ in the form
$$\Gamma_\gamma^{\alpha\beta}(t) = \eta^{\alpha\epsilon}
\partial_\epsilon\partial_\gamma f^\beta (t)
\eqno(2.27)$$
for a vector field $f^\beta (t)$. The equation (1.8) (or,
equivalently, (1.19)) for the metric $g^{\alpha\beta}(t)$ and the
connection (2.27) reads
$$g^{\alpha\sigma}\Gamma_\sigma^{\beta\gamma} = g^{\beta\sigma}
\Gamma_\sigma^{\alpha\gamma}.$$
For $\alpha = n$ because of Lemma 2.2 this gives
$$\sum_\sigma d_\sigma t^\sigma \eta^{\beta\epsilon}
\partial_\sigma\partial_\epsilon f^\gamma = (d_\gamma -1)g^{\beta\gamma}.
$$
Applying to the l.h.s. the Euler identity (here deg $\partial_\epsilon
f^\gamma = d_\gamma - d_\epsilon +h$) we obtain
$$(d_\gamma - 1) g^{\beta\gamma} =
\sum_\epsilon \eta^{\beta\epsilon} (d_\gamma - d_\epsilon +h)
\partial_\epsilon f^\gamma =
(d_\gamma + d_\beta - 2)\eta^{\beta\epsilon}\partial_\epsilon f^\gamma.
\eqno(2.28a)$$
{}From this one obtains the symmetry
$${\eta^{\beta\epsilon}\partial_\epsilon f^\gamma \over d_\gamma -1} =
{\eta^{\gamma\epsilon}\partial_\epsilon f^\beta \over d_\beta -1}.
$$
Let us denote
$${f^\gamma\over d_\gamma -1} =: {F^\gamma\over h} .
\eqno(2.28b)$$
We obtain
$$\eta^{\beta\epsilon}\partial_\epsilon F^\gamma =
\eta^{\gamma\epsilon}\partial_\epsilon F^\beta .$$
Hence a function $F(t)$ exists such that
$$F^\alpha = \eta^{\alpha\epsilon} \partial_\epsilon F.
\eqno(2.28c)$$
It is clear that $F(t)$ is a quasihomogeneous polynomial of
the degree $2h+2$. From the formula (2.28) one immediately obtains (2.25).

Let us prove now that the coefficients (2.26a) satisfy the associativity
condition. It is more convenient to work with the dual
structure constants
$$c_\gamma^{\alpha\beta}(t) = \eta^{\alpha\lambda}\eta^{\beta\mu}
\partial_\lambda\partial_\mu\partial_\gamma F.$$
Because of (2.27), (2.28) one has
$$\Gamma_\gamma^{\alpha\beta} = {d_\beta -1\over h} c_\gamma^{\alpha\beta}.$$
Substituting this in (1.18) we obtain associativity. Finaly,
for $\alpha = n$ the formulae (2.22), (2.23) imply
$$c_\beta^{n\alpha} = h \delta_\beta^\alpha .$$
Since $\eta^{1n} = h$, the vector (2.26b) is the unity of the algebra.
Lemma is proved.
\medskip
Proof of Theorem 1.

Existence of a Frobenius structure on the space of orbits satisfying
the conditions of Theorem 1 follows from Main lemma. We are now to prove
uniqueness.
Let us consider a polynomial Frobenius structure
on $M$ with the charges and dimension (3)
and with the Saito invariant metric. In the Saito flat coordinates we have
$$dt^\alpha\cdot dt^\beta = \eta^{\alpha\lambda} \eta^{\beta\mu}
\partial_\lambda \partial_\mu \partial_\gamma F(t)dt^\gamma .$$
The l.h.s. of (4) reads
$$i_v (dt^\alpha \cdot dt^\beta ) = \sum_\gamma d_\gamma t^\gamma
\eta_{\alpha\lambda} \eta^{\beta\mu}
\partial_\lambda \partial_\mu \partial_\gamma F(t)
= (d_\alpha + d_\beta - 2) \eta_{\alpha\lambda} \eta^{\beta\mu}
\partial_\lambda \partial_\mu  F(t) .$$
This should be equal to $h<dt^\alpha , dt^\beta >^*$. So the function
$F(t)$ should satisfy (2.25). It is determined uniquely by this equation
up to terms quadratic in $t^\alpha$. Such an ambiguity does not affect
the Frobenius structure. Theorem is proved.
\medskip
An algebraic remark: let $T$
be a $n$-dimensional space and $U: T\to T$ an endomorphism (linear
operator). Let
$$P_U(u) := {\rm det}\, (U-u\cdot 1) $$
be the characteristic polynomial of $U$. We say that the endomorphism $U$
is semisimple if all the $n$ roots of the characteristic polynomial are
simple. For a semisimple endomorphism there exists a cyclic vector
$e\in T$ such that
$$T = {\rm span}\, (e, Ue, ..., U^{n-1}e).$$
The map
$${\bf C}[u]/(P_U(u)) \to T, ~~u^k\mapsto U^ke, ~k=0, 1, ..., n-1
\eqno(2.29)$$
is an isomorphism of linear spaces.
\medskip
Let us fix a point $x\in M$. We define a linear operator
$$U = (U_j^i(x)): T_xM \to T_xM
\eqno(2.30)$$
(being also an operator on the cotangent bundle)
taking the ratio of the quadratic forms $g^{ij}$ and $\eta^{ij}$
$$(U\omega_1,\omega_2)^* = <\omega_1,\omega_2>^*
\eqno(2.31)$$
or, equivalently,
$$U_j^i(x) := \eta_{js}(x)g^{si}(x).
\eqno(2.32)$$
\medskip
{\bf Lemma 2.3.} {\it The characteristic polynomial of the operator
$U(x)$ is given up to a nonzero factor $c^{-1}$ (2.15) by the formula (5).}

Proof. We have
$$P(u;x^1, \dots , x^n) :=
{\rm det}(U - u\cdot 1) = {\rm det}(\eta_{js}){\rm det}
(g^{si} - u\eta^{si}) = $$
$$c^{-1}{\rm det}(g^{si}(x^1-u, x^2,
\dots , x^n) = c^{-1}D(x^1 - u, x^2, \dots , x^n).$$
Lemma is proved.
\medskip
{\bf Corollary 2.5.} {\it The operator $U(x)$ is semisimple at a generic
point $x\in M$.}

Proof. Let us prove that the discriminant $D_0(x^1, \dots , x^n)$ of the
characteristic polynomial $P(u;x^1, \dots , x^n)$ does not vanish identicaly
on $M$. Let us fix a Weyl chamber $V_0 \subset V$ of the group $W$. On the
inner
part of $V_0$ the factorization map
$$V_0 \to M_{Re}$$
is a diffeomorphism. On the image of $V_0$ the discriminant $D(x)$ is
positive. It vanishes on the images of the $n$ walls of the Weyl chamber:
$$D(x)_{i-{\rm th ~ wall}} = 0,~~ i=1, \dots , n.
\eqno(2.33)$$
On the inner part of the $i$-th wall (where the surface
(2.33) is regular) the equation (2.33) can be solved for $x^1$:
$$x^1 = x^1_i(x^2, \dots , x^n).
\eqno(2.34)$$
Indeed, on the inner part
$$(\partial_1D(x))_{i-{\rm th ~ wall}} \neq 0.$$
This holds since the polynomial
$D(x)$ has simple zeroes at the generic point of the discriminant of $W$
(see, e.g., [2]) .

Note that the functions (2.34) are the roots of the equation $D(x) = 0$
as the equation in the unknown $x^1$. It follows from above that this equation
has simple roots for generic $x^2$, ..., $x^n$. The roots of the characteristic
equation
$$D(x^1-u, x^2, \dots , x^n) = 0$$
are therefore
$$u_i = x^1 - x^1_i(x^2, \dots , x^n), ~~i=1, \dots , n.
\eqno(2.35)$$
Genericaly these are distinct. Lemma is proved.
\medskip
{\bf Lemma 2.4.} {The operator $U$ on the tangent planes $T_xM$ coincides
with the operator of multiplication by the Euler
vector field $v = {1\over h}E$.}

Proof. We check the statement of the lemma in the Saito flat coordinates:
$$\sum_\sigma {d_\sigma \over h} t^\sigma c_{\sigma\beta}^\alpha
= {h-d_\beta +d_\alpha\over h}\eta^{\alpha\epsilon}
\partial_\epsilon \partial_\beta F
=$$
$$ \sum_\lambda {d_\lambda + d_\alpha - 2 \over h}
\eta_{\beta\lambda}\eta^{\alpha\epsilon} \eta^{\lambda\mu}
\partial_\epsilon \partial_\mu F = \eta_{\beta\lambda}g^{\alpha\lambda}
= U_\beta^\alpha .$$
Lemma is proved.
\medskip
Proof of Theorem 2.

Because of Lemmas 2.3, 2.4  the vector fields
$$e, ~v, ~v^2, \dots , v^{n-1}
\eqno(2.36)$$
genericaly are linear independent on $M$. It is easy to see
that these are polynomial vector fields on $M$. Hence $e$ is a cyclic vector
for the endomorphism $U$ acting on $Der \, R$.
So in generic
point $x\in M$ the map (6a) is an isomorphism of Frobenius
algebras
$${\bf C}[u]/(P(u;x)) \to T_xM.$$
This proves Theorem 2.
\medskip
{\bf Remark 1.} The Euclidean metric (2.7) also defines an invariant
inner product for the Frobenius algebras (on the cotangent planes $T_*M$).
It can be shown also that the trilinear form
$$<\omega_1\cdot \omega_2, \omega_3>^*$$
can be represented (localy, outside the discriminant $Discr\, W$)
in the form
$$(\hat\nabla^i \hat\nabla^j \hat\nabla^k \hat F(x))
\partial_i \otimes \partial_j \otimes \partial_k$$
for some function $\hat F(x)$. Here $\hat\nabla$ is the Gauss-Manin
connection (i.e. the Levi-Civit\`a connection for the metric (2.7)).
The unity $dt^n/h$ of the Frobenius algebra on $T_*M$ is not
covariantly constant w.r.t. the Gauss-Manin connection.
\medskip
{\bf Remark 2.} The vector fields
$$l^i := g^{is}(x)\partial_s,~~i = 1, \dots , n
\eqno(2.37)$$
form a basis of the $R$-module Der$_R(-\log (D(x))$ of the vector
fields on $M$ tangent to the discriminant [2]. By the definition,
a vector field $u\in$ Der$_R(-\log (D(x))$ {\it iff}
$$uD(x) = p(x)D(x)$$
for a polynomial $p(x)\in R$. The basis (2.37) of
Der$_R(-\log (D(x))$ depends on the choice of coordinates
on $M$. In the Saito flat coordinates commutators of the basic vector
fields can be calculated via the structure constants of the Frobenius
algebra on $T_*M$. The following formula holds:
$$[l^\alpha , l^\beta ] = {d_\beta - d_\alpha \over h}
c^{\alpha\beta}_\epsilon l^\epsilon .
\eqno(2.38)$$
This can be proved using (2.25).
\medskip
{\bf Remark 3.} The eigenvalues $u_1(x)$, ..., $u_n(x)$ of the endomorphism
$U(x)$ can be chosen as new local coordinates near a generic point $x\in M$
(such that $D_0(x)\neq 0$). As it follows from [20, 22] these are {\it
canonical
coordinates} on the Frobenius manifold $M$: by the definition, this means
that the law of multiplication of the coordinate vector fields has the form
$$\partial_i \cdot \partial_j = \delta_{ij}\partial_i
\eqno(2.39)$$
$$\partial_i = {\partial\over \partial u_i}.$$
In these coordinates the Saito metric $(~,~)$ is given by a diagonal
Egoroff metric (see [20] for the definition)
$$\eta_{ij} (u) = \eta_{\alpha\beta}{\partial t^\alpha\over \partial u_i}
{\partial t^\beta\over \partial u_i}\delta_{ij}.
\eqno(2.40)$$
The Euclidean metric $<~,~>$ outside of the discriminant $u_1\dots u_n = 0$
in these coordinates is written as another diagonal Egoroff metric with the
diagonal entries $\eta_{ii}(u)/u_i$. The unity vector field has the form
$$e = \sum_{i=1}^n \partial_i
\eqno(2.41)$$
and the Euler vector field
$${1\over h} E = v = \sum_{i=1}^n u_i\partial_i .
\eqno(2.42)$$
I recall that, according to the theory of [20] the metric (2.40) satisfies
the Darboux-Egoroff equations
$$\partial_k\gamma_{ij} = \gamma_{ik} \gamma_{kj}, ~~
i, j, k ~{\rm are ~ distinct},
\eqno(2.43a)$$
$$\sum_{k=1}^n \partial_k \gamma_{ij} = 0
\eqno(2.43b)$$
$$\sum_{k=1}^n u_k\partial_k \gamma_{ij} = -\gamma_{ij}
\eqno(2.43c)$$
where the rotation coefficients $\gamma_{ij}(u) = \gamma_{ji}(u)$ are defined
by the formula
$$\gamma_{ij}(u) := {\partial \sqrt{\eta_{jj}(u)}\over
\sqrt{\eta_{ii}(u)}}, ~~i\neq j.
\eqno(2.44)$$
The system (2.43) is empty for $n=1$; it is linear for $n=2$. For the first
nontrivial case $n=3$ it can be reduced to a particular case of the
Painlev\'e-VI equation [27] using the first integral
$$(u_1 - u_2)^2 \gamma^2_{12} +
(u_1 - u_3)^2 \gamma^2_{13} +
(u_2 - u_3)^2 \gamma^2_{23} = R^2.
\eqno(2.45)$$
For any $n\geq 3$ the system (2.43) can be reduced to a system of ordinary
differential equations. It
coincides with the equations of
isomonodromy deformations of a certain linear differential operator
with rational coefficients [20, 22]. Thus the eqs. (2.43) can be called a
high-order analogue of the Painlev\'e-VI. The constructions of the
present paper for the groups $A_3$, $B_3$, $H_3$ specify three
distinguished solutions of the correspondent Painlev\'e-VI eqs..
The function $F(t)$ for these groups has the form
$$F_{A_3} = {t_1^2t_3+t_1t_2^2\over 2} + {t_2^2t_3^2\over 4} + {t_3^5\over 60}
\eqno(2.46)$$
$$F_{B_3} = {t_1^2t_3+t_1t_2^2\over 2} + {t_2^3t_3\over 6} + {t_2^2t_3^3\over
6}
+{t_3^7\over 210}
\eqno(2.47)$$
$$F_{H_3} = {t_1^2t_3+t_1t_2^2\over 2} + {t_2^3t_3^2\over 6}
+{t_2^2t_3^5\over 20} +{t_3^{11}\over 3960}.
\eqno(2.48)$$
The correspondent constants $R$ in (2.45) equal $1/4$, $1/3$ and $2/5$ resp.
\medskip
\medskip
{\bf Concluding remarks.}
\medskip
1. The results of this paper can be generalised for the case where $W$ is the
Weyl group of an extended affine root system of codimension 1 (see the
definition in [39]). In this case the Frobenius structure will be
polynomial in all the coordinates but one and it will be a modular form in
this exceptional coordinate. The solutions of WDVV
of [32, 46] are just of this type. We consider the orbit spaces of
these groups in a subsequent publication.

\medskip
2. The two metrics on the space of orbits of the group $A_n$ are closely
related to the two hamiltonian structures of the $n$KdV hierarchy (see
[18 - 20, 22]). The Saito metric is obtained by the semiclassical limit of
[24, 25] from the first Gelfand-Dickey Poisson bracket of $n$KdV, and the
Euclidean metric is obtained by the same semiclassical limit from the second
Gelfand-Dickey Poisson bracket. The Saito and the Euclidean coordinates
on the orbit space are the Casimirs for the corresponding Poisson brackets.
The factorization map $V\to M=V/W$ is the semiclassical limit of the Miura
transformation.
Probably, the semiclassical limit of the
bi-hamiltonian structure of the $D-E$ Drinfeld-Sokolov hierarchies [17] give
the
two flat metrics on the orbit spaces of the groups $D_n$ and $E_6$, $E_7$,
$E_8$ resp. But this should be checked.

It is still an open question if it is possible to relate integrable
hierarchies to the Coxeter groups not of $A-D-E$ series. A partial
answer to this question is given in [20, 22]: the
unknown integrable hierarchies for any Frobenius manifold are constructed
in a semiclassical (i.e., in the dispersionless) approximation.
\medskip
3. A closely related question: what is the algebraic-geometrical description
of the TFT models related to the polynomial solutions of WDVV constructed in
this paper? For $A-D-E$ groups the correspondent TFT models are the topological
minimal models of [15]. For other Coxeter groups the TFT can be
constructed as equivariant topological Landau-Ginsburg models
using the results of [44, 47] for $W\neq H_4$ (the singularity theory
related to $H_4$ was partialy developed in [35, 40]). For the group $A_n$
a nice algebraic-geometrical reformulation of the correspondent TFT as the
intersection theory on a certain covering over the moduli space of stable
algebraic curves, was proposed in [50, 51] (for the topological gravity
$W=A_1$ this
conjecture was proved by M.Kontsevich [29, 30]). What are the moduli spaces
whose intersection theories are encoded by the orbit spaces of other
Coxeter groups? Note that a part of these intersection numbers should
coincide with the coefficients of the polynomials $F(t)$ (these are rational
but not integer numbers since the moduli spaces are not manifolds but
orbifolds).
\medskip
\medskip
{\bf Acknowledgements.}
\medskip
I am grateful to V.I.Arnol'd for fruitful discussions.

\vfill\eject

{\bf Appendix. Algebraic version of the definition of polynomial
Frobenius manifold.}
\medskip
Let $k$ be a field of the characteristic $\neq 2$ and
$$R := k[x^1, \dots ,x^n]
\eqno(A.1)$$
be the ring of polynomials with the coefficients in $k$. By $Der\, R$
we denote the $R$-module of $k$-derivations of $R$. This is a free
$R$-module with the basis
$$\partial_i := {\partial \over \partial x^i}, ~~ i = 1, \dots , n.$$
A map
$$Der\, R \times R \to R$$
is defined by the formula
$$(u=u^i\partial_i, ~p) \mapsto up := u^i\partial_ip.
\eqno(A.2)$$
A $R$-bilinear symmetric inner product
$$Der\, R \times Der\, R \to R$$
$$u, ~v \mapsto (u,v)\in R
\eqno(A.3)$$
is called nondegenerate if from the equations
$$(u,v) = 0 ~~{\rm for ~ any ~} v\in Der\, R$$
it follows that $u=0$.

As it was mentioned in Introduction, a polynomial
Frobenius manifold is a structure of Frobenius $R$-algebra on
$Der\, R$ satisfying certain conditions. We obtain here these conditions
by reformulating the Definition 1 in a pure algebraic way.

The first standard step is to reformulate the notion of the Levi-Civit\`a
connection. By the definition, this is a map
$$Der\, R\times Der\, R \to Der\, R$$
$$u, ~v \mapsto \nabla_uv
\eqno(A.4)$$
$R$-linear in the first argument and satisfying the Leibnitz rule in the
second one
$$\nabla_u(pv) = p\nabla_uv + (up) v
\eqno(A.5)$$
uniquely specified by the equations
$$u(v,w) = (\nabla_uv, w) + (v, \nabla_uw)
\eqno(A.6a)$$
$$\nabla_uv - \nabla_vu = [u,v]
\eqno(A.6b)$$
(the commutator of the derivations). Equivalently, it can be determined
from the equation
$$<\nabla_uv,w>=
{1\over 2}[u<v,w>+v<w,u>-w<u,v>+
$$
$$<[u,v],w>+<[w,u],v>+<[w,v],u>]
\eqno(A.6c)$$
for arbitrary $u$, $v$, $w\in Der\, R$.

Now the assumptions 1 - 3 of Definition 1 for the Frobenius $R$-algebra
$Der\, R$ can be reformulated as follows:

1. For any $u$, $v$, $w$ the following identity holds
$$(\nabla_u\nabla_v - \nabla_v\nabla_u - \nabla_{[u,v]})w= 0.$$

2. For the unity $e\in Der\, R$ and for arbitrary $u\in Der\, R$
$$\nabla_ue = 0.$$

3. The identity
$$\nabla_u(v\cdot w)-\nabla_v(u\cdot w)
+u\cdot\nabla_vw-v\cdot\nabla_uw=[u,v]\cdot w
\eqno(A.7)$$
holds for any three derivations fields $u$, $v$, $w$.

To reformulate the assumption 4 of Definition 1 let us assume that
$Der\, R$ is a graded algebra over a graded ring $R$ with a graded
invariant inner product $(~,~)$. That means that two gradings
$\degg$ and $\deg'$ are defined
on $R$ and on $Der\, R$ resp., i.e. real numbers
$$P_i := \degg x^i, ~~Q_i := \deg' \partial_i
\eqno(A.8)$$
are assigned to the generators $x^1$, ..., $x^n$ and to the basic
derivations $\partial_1$, ..., $\partial_n$ resp. By the definition,
the degree of a monomial
$$p = (x^1)^{m_1}\dots (x^n)^{m_n}$$
equals
$$\degg p := m_1P_1 + \dots +m_nP_n.$$
Homogeneous elements of $Der \, R$ are defined by the assumption that
the operators $p\mapsto up$ shifts the grading in $R$ to $\deg' u - Q_0$
for a constant $Q_0$, i.e.
$$\degg (up) = \deg' u + \degg p - Q_0.
\eqno(A.9)$$
The $R$-algebra structure on $Der\, R$ should be consistent with the grading,
i.e. for any homogeneous elements $p,~q$
of $R$ and $u,~v$ of $Der\, R$ the following formulae
hold:
$$\deg' (pu) = \deg' u + \degg p
\eqno(A.10)$$
$$\degg (pq)  = \degg p + \degg q
\eqno(A.11)$$
$$\deg' (u\cdot v) = \deg' u + \deg' v.
\eqno(A.12)$$
The invariant inner product $(~,~)$ should be graded of a degree $D$, i.e.
$$(u,v) = 0 ~~{\rm if }~~ \deg' u + \deg' v \neq D
\eqno(A.13)$$
for arbitrary homogeneous $u,~v \in Der\, R$. Note that the Euler
vector field is  homogeneous of the degree $Q_0$. We consider only the case
$Q_0 \neq 0$.

The numbers $P_i$, $Q_i$, $Q_0$,
$D$ are defined up to rescaling. One can normalise
these in such a way that $Q_0 = 1$.
Then we have
$$Q_i :=q_i, ~P_i = 1-q_i ,~ D=d$$
in the notations of Introduction.

The constructions of this paper give such an algebraic structure for
$k = {\bf Q}$.

\vfill\eject

{\bf References.}
\medskip
\item{1.} Arnol'd V.I., Normal forms of functions close to degenerate critical
points. The Weyl groups $A_k$, $D_k$, $E_k$, and Lagrangian singularities,
{\sl Functional Anal.} {\bf 6} (1972) 3 - 25.
\medskip
\item{2.} Arnol'd V.I., Wave front evolution and equivariant Morse lemma,
{\sl Comm. Pure Appl. Math.} {\bf 29} (1976) 557 - 582.
\medskip
\item{3.} Arnol'd V.I., Indices of singular points of 1-forms on a manifold
with boundary, convolution of invariants of reflection groups, and singular
projections of smooth surfaces, {\sl Russ. Math. Surv.} {\bf 34} (1979)
1 - 42.
\medskip
\item{4.} Arnol'd V.I., Gusein-Zade S.M., and Varchenko A.N., Singularities
of Differentiable Maps, volumes I, II, Birkh\"auser, Boston-Basel-Berlin,
1988.
\medskip
\item{5.} Arnol'd V.I., Singularities of Caustics and Wave Fronts,
Kluwer Acad. Publ., Dordrecht - Boston - London, 1990.
\medskip
\item{6.} Aspinwall P.S., Morrison D.R., Topological field theory and
rational curves, {\sl Comm. Math. Phys.} {\bf 151} (1993) 245 - 262.
\medskip
\item{7.} Atiyah M.F., Topological quantum field theories, {\sl Publ. Math.
I.H.E.S.} {\bf 68} (1988) 175.
\medskip
\item{8.} Blok B. and Varchenko A., Topological conformal field theories
and the flat coordinates, {\sl Int. J. Mod. Phys.} {\bf A7} (1992) 1467.
\medskip
\item{9.} Bourbaki N., Groupes et Alg\`ebres de Lie, Chapitres 4, 5 et 6,
Masson, Paris-New York-Barcelone-Milan-Mexico-Rio de Janeiro, 1981.
\medskip
\item{10.} Brieskorn E. Singular elements of semisimple algebraic groups,
In: Actes Congres Int. Math., {\bf 2}, Nice (1970), 279 - 284.
\medskip
\item{11.} Cecotti S. and Vafa C., {\sl Nucl. Phys.} {\bf B367} (1991) 359.
\medskip
\item{12.} Cecotti S. and  Vafa C., On classification of $N=2$ supersymmetric
theories, Preprint HUTP-92/A064 and SISSA-203/92/EP, December 1992.
\medskip
\item{13.} Coxeter H.S.M., Discrete groups generated by reflections,
{\sl Ann. Math.} {\bf 35} (1934) 588 - 621.
\medskip
\item{14.} Dijkgraaf R. and Witten E., {\sl Nucl. Phys.} {\bf B 342} (1990) 486
\medskip
\item{15.} Dijkgraaf R., E.Verlinde, and H.Verlinde, {\sl Nucl. Phys.}
{\bf B 352} (1991) 59;
\item{} Notes on topological string theory and 2D quantum gravity,
Preprint PUPT-1217, IASSNS-HEP-90/80, November 1990.
\medskip
\item{16.} Dijkgraaf R., Intersection theory, integrable hierarchies and
topological field theory, Preprint IASSNS-HEP-91/91, December 1991.
\medskip
\item{17.} Drinfel'd V.G. and Sokolov V.V., {\sl J. Sov. Math.} {\bf 30} (1985)
1975.
\medskip
\item{18.} Dubrovin B., Differential geometry of moduli spaces and its
application to soliton equations and to topological field theory,
Preprint No.117,  Scuola Normale Superiore, Pisa (1991).
\medskip
\item{19.} Dubrovin B., Hamiltonian formalism of Whitham-type hierarchies
and topological Landau - Ginsburg models, {\sl Comm. Math. Phys.}
{\bf 145} (1992) 195 - 207.
\medskip
\item{20.} Dubrovin B., Integrable systems in topological field theory,
{\sl Nucl. Phys.} {\bf B 379} (1992) 627 - 689.
\medskip
\item{21.} Dubrovin B., Geometry and integrability of
topological-antitopological fusion, Pre\- print INFN-8/92-DSF, to appear
in {\sl Comm. Math. Phys.}
\medskip
\item{22.} Dubrovin B., Integrable systems and classification of
2-dimensional topological field theories, Preprint SISSA 162/92/FM,
September 1992, to appear in \lq\lq Integrable Systems", Proceedings
of Luminy 1991 conference dedicated to the memory of J.-L. Verdier.
\medskip
\item{23.} Dubrovin B., Topological conformal field theory from the point
of view of integrable systems, Preprint SISSA 12/93/FM, January 1993,
to appear in Proceedings of 1992 Como workshop \lq\lq Quantum Integrable
Systems".
\medskip
\item{24.} Dubrovin B. and Novikov S.P., The Hamiltonian formalism
of one-dimensional systems of the hydrodynamic type and
the Bogoliubov - Whitham averaging method,{\sl Sov. Math. Doklady}
{\bf 27} (1983) 665 - 669.
\medskip
\item{25.} Dubrovin B. and Novikov S.P., Hydrodynamics of weakly deformed
soliton lattices. Differential geometry and Hamiltonian theory,
{\sl Russ. Math. Surv.} {\bf 44:6} (1989) 35 - 124.
\medskip
\item{26.} Dubrovin B., Novikov S.P., and Fomenko A.T., Modern Geometry,
Parts 1 - 3, Springer Verlag.
\medskip
\item{27.} Fokas A.S., Leo R.A., Martina L., and Soliani G., {\sl Phys. Lett.}
{\bf A115} (1986) 329.
\medskip
\item{28.} Givental A.B., Convolution of invariants of groups generated
by reflections, and connections with simple singularities of functions,
{\sl Funct. Anal.} {\bf 14}  (1980) 81 - 89.
\medskip
\item{29.} Kontsevich M.,  {\sl Funct. Anal.} {\bf 25} (1991) 50.
\medskip
\item{30.} Kontsevich M., {\sl Comm. Math. Phys.} {\bf 147} (1992) 1.
\medskip
\item{31.} Looijenga E., A period mapping for certain semiuniversal
deformations, {\sl Compos. Math.} {\bf 30} (1975) 299 - 316.
\medskip
\item{32.} Maassarani Z., {\sl Phys. Lett.} {\bf 273B} (1992) 457.
\medskip
\item{33.} Magri F., {\sl J. Math. Phys.} {\bf 19} (1978) 1156.
\medskip
\item{34.} Procesi C. and Schwarz G., Inequalities defining orbit spaces,
{\sl Invent. Math.} {\bf 81} (1985) 539 - 554.
\medskip
\item{35.} Roberts R.M. and Zakalyukin V.M., Symmetric wavefronts, caustic
and Coxeter groups, to appear in Proceedings of Workshop in the Theory of
Singularities, Trieste 1991.
\medskip
\item{36.} Saito K., On a linear structure of a quotient variety by a finite
reflection group, Preprint RIMS-288 (1979).
\medskip
\item{37.} Saito K., Yano T., and Sekeguchi J., On a certain generator system
of the ring of invariants of a finite reflection group,
{\sl Comm. in Algebra} {\bf 8(4)} (1980) 373 - 408.
\medskip
\item{38.} Saito K., Period mapping associated to a primitive form,
{\sl Publ. RIMS} {\bf 19} (1983) 1231 - 1264.
\medskip
\item{39.} Saito K., Extended affine root systems II (flat invariants),
{\sl Publ. RIMS} {\bf 26} (1990) 15 - 78.
\medskip
\item{40.} Shcherbak O.P., Wavefronts and reflection groups, {\sl Russ. Math.
Surv.} {\bf 43:3} (1988) 149 - 194.
\medskip
\item{41.} Slodowy P., Einfache Singularitaten und Einfache Algebraische
Gruppen,
Preprint, Regensburger Mathematische Schriften {\bf 2}, Univ. Regensburg
(1978).
\medskip
\item{42.} Vafa C., {\sl Mod. Phys. Let.} {\bf A4} (1989) 1169.
\medskip
\item{43.} Vafa C., Private communication, September 1992.
\medskip
\item{44.} Varchenko A.N. and Chmutov S.V., Finite irreducible groups,
generated
by reflections, are monodromy groups of suitable singularities,
{\sl Func. Anal.} {\bf 18} (1984)
171 - 183.
\medskip
\item{45.} Varchenko A.N. and Givental A.B., Mapping of periods and
intersection form, {\sl Funct. Anal.} {\bf 16} (1982) 83 - 93.
\medskip
\item{46.} Verlinde E. and Warner N., {\sl Phys. Lett.} {\bf 269B} (1991) 96.
\medskip
\item{47.} Wall C.T.C., A note on symmetry of singularities, {\sl Bull. London
Math. Soc.} {\bf 12} (1980) 169 - 175;
\item{} A second note on symmetry of singularities, {\sl ibid.}, 347 - 354.
\medskip
\item{48.} Witten E., {\sl Comm. Math. Phys.} {\bf 117} (1988) 353;
\item{} {\sl ibid.}, {\bf 118} (1988) 411.
\medskip
\item{49.} Witten E., {\sl Nucl. Phys.} {\bf B 340} (1990) 281.
\medskip
\item{50.} Witten E., {\sl Surv. Diff. Geom.} {\bf 1}  (1991) 243.
\medskip
\item{51.} Witten E., Algebraic geometry associated with matrix models of
two-dimensional gravity, Preprint IASSNS-HEP-91/74.
\medskip
\item{52.} Witten E., Lectures on mirror symmetry, In: Proceedings MSRI
Conference
on mirror symmetry, March 1991, Berkeley.
\medskip

\vfill\eject\end